\documentclass[
aps,
prd,
preprint,
superscriptaddress,
titlepage,
a4paper,
showkeys,
amsfonts,
amssymb,
amsmath,
]{revtex4-2}

\usepackage{graphicx}
\usepackage{amsmath}
\usepackage{amsfonts}
\usepackage{amssymb}
\usepackage{xcolor}
\usepackage{xspace}
\usepackage[colorlinks=true, allcolors=blue]{hyperref}
\usepackage[capitalise]{cleveref}
\usepackage{threeparttable}
\usepackage{multirow}
\usepackage{enumerate}
\usepackage{subfig}
\usepackage{textgreek}
\usepackage{float}
\usepackage{threeparttable}
\usepackage{booktabs}

\begin{document}

\title{Design of a CsI(Tl) Calorimeter for Muonium-to-Antimuonium Conversion Experiment}

\thanks{Supported by the National Natural Science Foundation of China (No.12075326), the Natural Science Foundation of Guangzhou (No.2024A04J6243) and Fundamental Research Funds for the Central Universities (23xkjc017) in Sun Yat-sen University.}

\author{Siyuan Chen}
\affiliation{School of Physics, Sun Yat-sen University, Guangzhou 510275, China}
\author{Shihan Zhao}
\affiliation{School of Physics, Sun Yat-sen University, Guangzhou 510275, China}
\author{Weizhi Xiong}
\affiliation{Institute of Frontier and Interdisciplinary Science, Shandong University, Qingdao 266237, China}
\author{Ye Tian}
\affiliation{Institute of Modern Physics, Chinese Academy of Science, Lanzhou 730000, China}
\author{Hui Jiang}
\affiliation{Institute of High Energy Physics, Chinese Academy of Science, Beijing 100049, China}
\affiliation{University of Chinese Academy of Sciences, Beijing 100049, China}
\author{Jiacheng Ling}
\affiliation{School of Physics, Sun Yat-sen University, Guangzhou 510275, China}
\author{Shishe Wang}
\affiliation{School of Physics, Sun Yat-sen University, Guangzhou 510275, China}
\author{Jian Tang}
\email[Corresponding author, ]{tangjian5@mail.sysu.edu.cn}
\affiliation{School of Physics, Sun Yat-sen University, Guangzhou 510275, China}

\date{\today}

\begin{abstract}
    The Muonium-to-Antimuonium Conversion Experiment (MACE) is proposed to search for charged lepton flavor violation and increase the sensitivity by more than two orders of magnitude compared to the MACS experiment at PSI in 1999. A clear signature of this conversion is the positron produced from antimuonium decay. This paper presents a near-$4\pi$-coverage calorimeter designed for MACE, which can provide an energy resolution of 10.8\% at 511 keV, and a signal efficiency of 78.3\% for annihilation $\gamma$-ray events. Detailed Monte-Carlo simulations using MACE offline software based on \textsc{Geant4} are performed for geometry optimization, coincidence system design, background estimation, and benchmark detector validation.
\end{abstract}


\maketitle
\clearpage

\section{Introduction} \label{sec:1}
In the Standard Model (SM), lepton flavor violation is strictly forbidden.
However, the discovery of neutrino oscillations has shown that neutral leptons can violate this conservation law, hinting at the existence of new physics beyond the Standard Model \cite{DayaBay2012,JUNO2015}.
Consequently, the Charged Lepton Flavor Violation (CLFV) is naturally expected within the framework of new physics. Muon, the lightest massive unstable charged lepton with a reasonably long lifetime, serves as an ideal probe in precision frontier experiments aimed at searching for CLFV processes.
Since the 1940s, several CLFV experiments with muon have been conducted, consistently reporting increasingly stringent upper limits over the decades.
Key channels of interest include $\mu^+ \rightarrow e^+\gamma$, $\mu^+ \rightarrow e^+e^-e^+$, and $\mu^-N \rightarrow e^-N$~\cite{Perrevoort2023}.
In recent years, notable experiments searching for these CLFV channels include MEG and its upgrade, MEG-II~\cite{Baldini2018}, Mu3e~\cite{Arndt2021}, COMET~\cite{Abramishvili2020}, and Mu2e~\cite{Bartoszek2014}.
However, the processes mentioned above violate lepton flavor by only one unit.

Muonium (M) is an exotic atom consisting of a bound state of $\mu^+e^-$, while its antiatom, antimuonium ($\bar{\text{M}}$), is composed of $\mu^-e^+$.
From the perspective of the Standard Model Effective Field Theory (SMEFT), the spontaneous M-to-$\bar{\text{M}}$ conversion ($\mu^+e^-\rightarrow\mu^-e^+$), which violates lepton flavor by two units instead of one, can probe distinct effective operators and thereby constrain a different parameter space~\cite{Fukuyama2022,Conlin:2022sga,Heeck2024}.
As a result, the key innovation in investigating M-to-$\bar{\text{M}}$ conversion is providing additional constraints on specific new physics models, distinct from other CLFV processes~\cite{Bernstein2013,Han2021,Afik2024}.
The most recent measured upper limit ($8.3\times10^{-11}$ at 90\%~C.L.) for the conversion probability in a 0.1~T magnetic field was reported by the MACS experiment at PSI in 1999~\cite{Willmann1999,Willmann2021}.
Another experiment focused on M-to-$\bar{\text{M}}$ conversion was proposed and conceptually designed at PRISM in 2003, as reported in~\cite{Aoki2003}.
There is also a proposal at J-PARC in Japan to search for this process using a novel scheme~\cite{Kawamura2021}.
But none of these have progressed to further studies.
Surprisingly, the limit of M-to-$\bar{\text{M}}$ conversion has remained unchallenged for two decades. There is an urgent need for a new-generation experiment.

\begin{figure}[htbp]
    \centering
    \includegraphics[width=.9\linewidth]{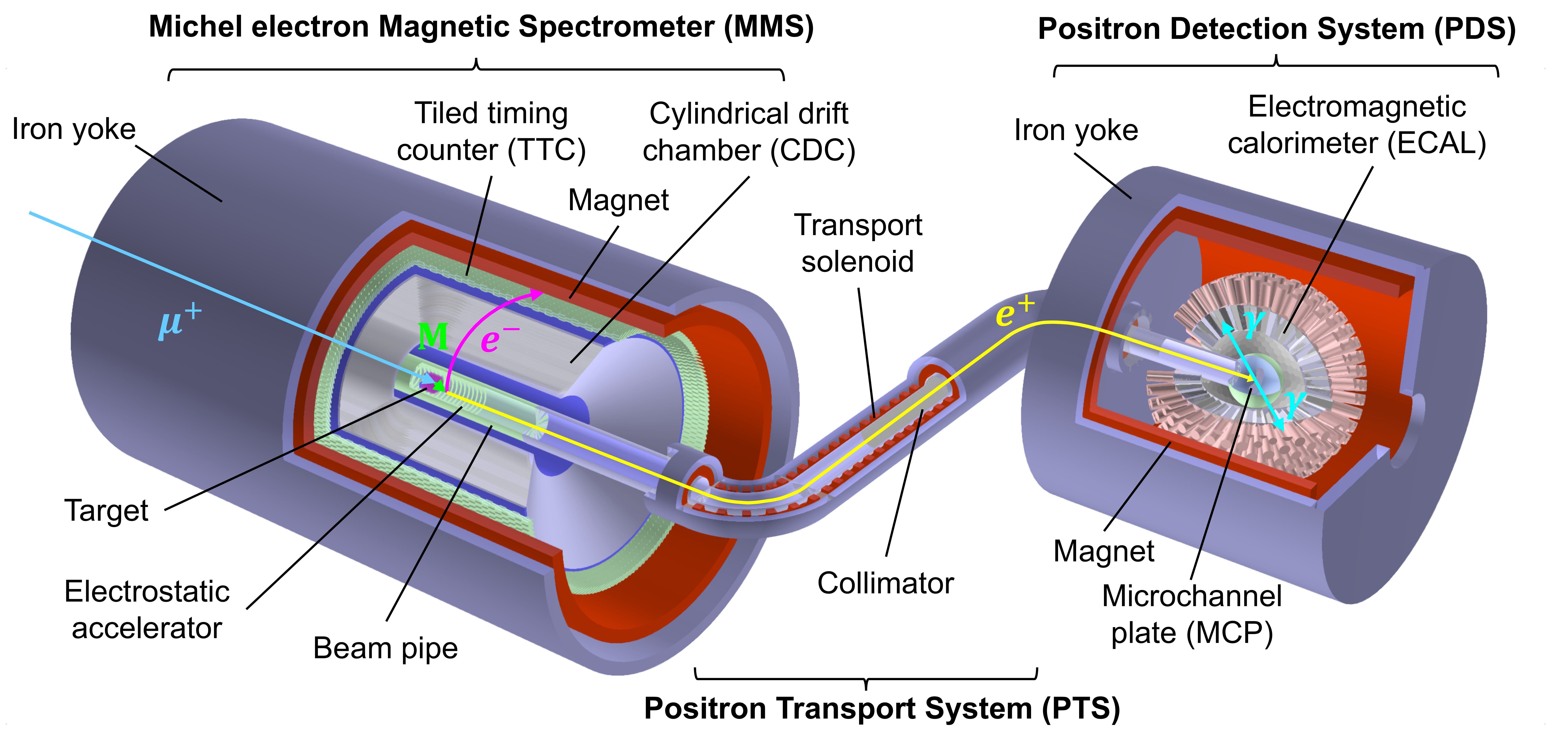}
    \caption{Schematic view of the MACE detector with a simulated M-to-$\bar{\text{M}}$ signal.}
    \label{fig:mace}
\end{figure}

In light of significant advancements in muon beamlines and particle detector techniques nowadays, the Muonium-to-Antimuonium Conversion Experiment (MACE) is proposed to search for M-to-$\bar{\text{M}}$ conversion and has the potential to improve the upper limit by three orders of magnitude~\cite{Bai2022,Bai:2024skk}.
With the construction of the China advanced NUclear physics research Facility (CNUF) and other local muon sources~\cite{Huizhou2020,Cai2024,Hong2024,Lv:2023wsc}, MACE is expected to be the first muon-CLFV experiment in China, offering unique perspectives compared to other related experiments, including those at colliders. The proposed MACE detector consists of a Michel electron Magnetic Spectrometer (MMS), a Positron Transport System (PTS), a Microchannel Plate (MCP) \cite{Miao2023,Peng2024}, and an Electromagnetic Calorimeter (ECAL) to probe the conversion signal through final state coincidence. The specifications of muon beamline in MACE are expected to produce a surface muon flux on target of $10^8~\mu^+/$s, with a repetition rate of 30–50 kHz and a pulse width duty cycle below 20\%.
A schematic of the MACE apparatus is shown in \cref{fig:mace}.

In the MACE concept, surface muons with a momentum of 26.3~MeV$/c$ are stopped at a production target where M is formed and then possibly diffuses into the vacuum and converts into $\bar{\text{M}}$.
The decay products of $\bar{\text{M}}$ include a Michel electron with an energy up to 52.8~MeV and a positron with atomic-shell energy ($\langle E_\text{k}\rangle=13.5$~eV).
The Michel electron will be reconstructed by MMS which is expected to achieve $10^{-4}$ level suppression of background in the ECAL end through preliminary simulation.
On the other hand, the atomic-shell positron will be accelerated by an electric field and guided through the PTS. The transverse position of the $e^+$ is conserved during the transportation to MCP, with a flight time full width at half maximum (FWHM) of 6.9~ns.
The cylindrical magnet with a field strength of 0.1 T outside the ECAL is used to constrain the positron beam spot after transportation.
When the MCP, with a timing resolution of 500 ps, is hit by a positron, the ECAL will be triggered and measure the produced two back-to-back annihilation $\gamma$-rays.
As a rare-process searching experiment, MACE requires high sensitivity for the annihilation $\gamma$-rays to discriminate the signal from backgrounds and retain as much event information as possible.
Thus, an ECAL with good energy resolution and high efficiency is essential for the design of MACE.

In modern particle physics experiments, the demand for high precision continues to evolve.
The calorimeter, which utilizes various active materials, is a crucial instrument used in most of the experiments for particle energy measurement up to date.
The geometry of the calorimeter is a key factor in signal efficiency \cite{Fabjan2003}.
Hermeticity, which refers to the coverage of the interaction vertex, is important for enhancing efficiency, particularly in rare-process search experiments.
Since the 1980s, various experiments have proposed near-$4\pi$-coverage spherical geometries for calorimeters based on these considerations.
One of the first $4\pi$ calorimeters is the Crystal Ball detector~\cite{Oreglia1982}.
It consists of 624 NaI(Tl) crystals and covers 94\% of $4\pi$ solid angle.
The subsequent PIBETA detector at PSI adopts a comparable design and is constructed using pentagonal, hexagonal, and trapezoidal polyhedron crystals, rather than triangular prism crystals used in the Crystal Ball, to achieve a higher light collection efficiency.
The PIBETA calorimeter has 240 pure CsI crystals and a 77\% coverage of $4\pi$ solid angle~\cite{Frlez2004}.
A next-generation experiment PIONEER at PSI is also investigating the similar design as an alternative using LYSO crystals~\cite{PIONEER:2022yag,Beesley:2024mts}.
In addition to the high-energy physics experiments, the $4\pi$ calorimeters are also widely used in nuclear physics experiments.
The Spin Spectrometer, constructed at the Oak Ridge National Laboratory, which is the contemporary of the Crystal Ball, was designed to measure the multiplicity of $\gamma$-rays in nuclear reactions~\cite{Jaeaeskelaeinen1983}.
It features 72 NaI(Tl) crystals, a 96.8\% solid angle coverage, with 7.7\%$-$9.2\% energy resolution at 662~keV.
Subsequently, the Detector for Advanced Neutron Capture Experiment (DANCE) at Los Alamos National Laboratory~\cite{Heil2001}, the Total Absorption Calorimeter (TAC) at CERN~\cite{Guerrero2009}, and the Gamma-ray Total Absorption Facility (GTAF) at CSNS~\cite{Zhang2017} have all utilized and refined this design.
The advantages of a $4\pi$ calorimeter include (1) near-4$\pi$ coverage for a high detection efficiency, (2) an isotropic spatial distribution and a good symmetry for precise reconstruction, (3) a small number of module types, (4) a simplified mechanical support scheme with self-supporting modules.
These features contribute to improving the precision of energy measurements in MACE.

The main purpose of this study is to propose a design for the MACE $4\pi$ calorimeter using the Monte-Carlo simulation method.
In this work, the geometry of ECAL is generated and optimized using the MACE offline software based on the \textsc{Geant4} toolkit~\cite{Agostinelli2003,Zhao2023}.
Subsequently, the energy resolution, signal efficiency, and background level of ECAL are estimated, providing a reference for the MACE conceptual design.
This paper is structured as follows:
\cref{sec:1} introduces the MACE calorimeter design and geometry optimization.
\cref{sec:2} presents the simulation results for both signal and background.
\cref{sec:3} describes the experiment setups of a benchmark detector, specifically the assembly used for an initial test of the CsI(Tl) crystal.
The paper concludes with a discussion of the conclusions and perspectives for future work in \cref{sec:4}.

\section{Design and Simulation of the MACE Calorimeter}\label{sec:2}
\subsection{Overview}
To meet the requirements of enhanced sensitivity (e.g., an excellent energy resolution, a high detection efficiency, and a good geometric acceptance),
we propose a near-$4\pi$-coverage polyhedral calorimeter design.
In this section, we present a design scheme for ECAL, including details on the selection of scintillation materials and photosensors.
This proposed design will serve as a basis for the simulation work discussed in the following sections.

The $4\pi$ ECAL composed of polyhedral crystals is geometrically referred to as the Goldberg polyhedron in mathematics, which was first introduced by Michel Goldberg in 1937~\cite{Goldberg1937}.
The Goldberg polyhedron consists of pentagons and hexagons with every three faces meeting at the same vertex.
It is similar in appearance to the sphere as an inpolyhedron and possesses an icosahedral rotational symmetry.
We generate the Goldberg polyhedron mesh using the following steps:

\begin{enumerate}[(1)]
    \item Generate a regular icosahedron mesh inscribed on the unit sphere;
    \item Apply the Loop subdivision algorithm~\cite{Loop1987} to the mesh and project new vertices to the sphere;
    \item Project the centroids of each face onto the sphere and connect them to form a Goldberg polyhedron mesh.
    \item Scale the mesh radially to the desired size.
\end{enumerate}

\begin{figure}[!t]
    \centering
    \subfloat[3D rendered image of the crystals.]{\includegraphics[width=0.7\linewidth]{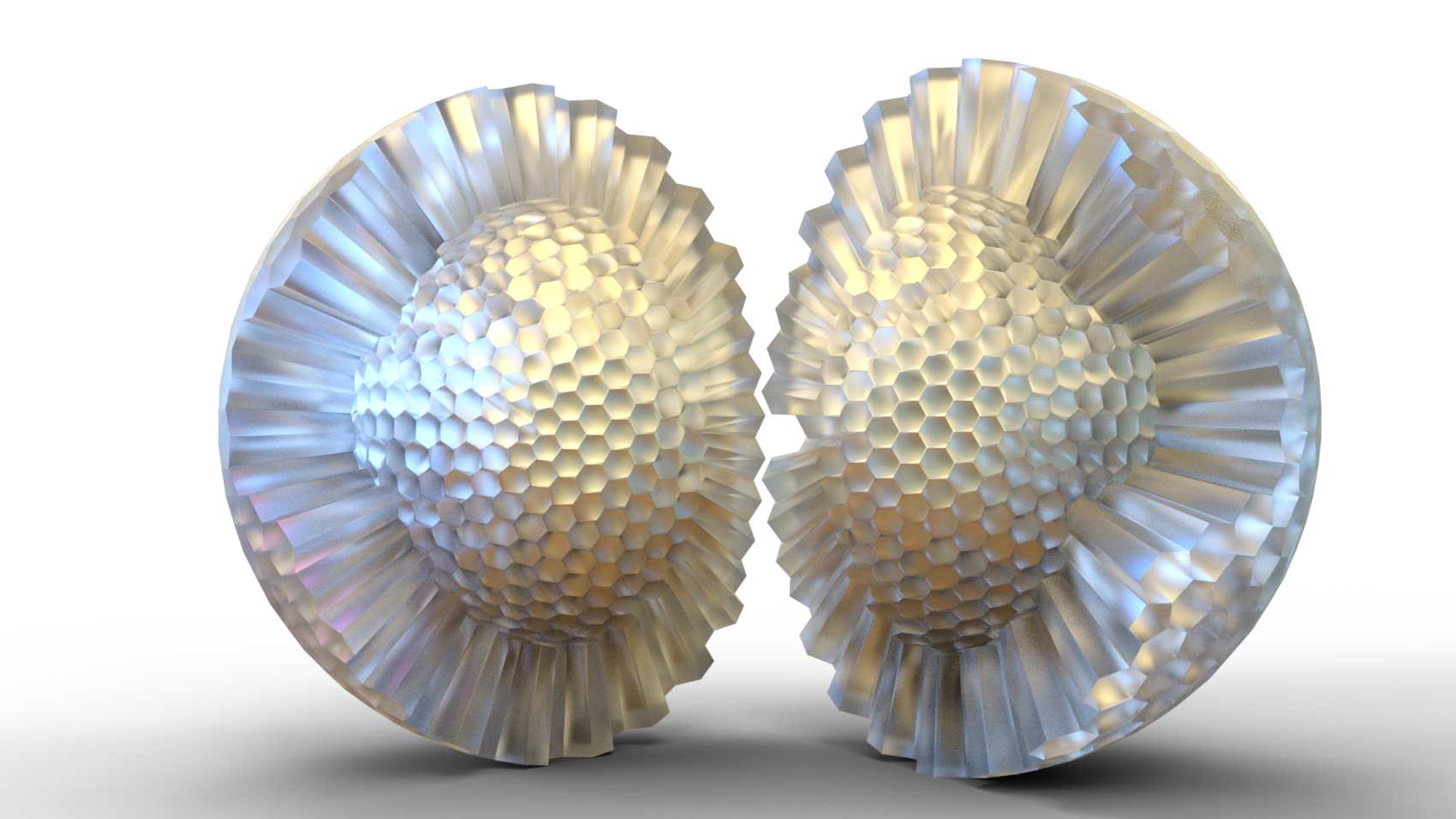}}
    \\
    \subfloat[An event display for ECAL.]{\includegraphics[width=0.7\linewidth]{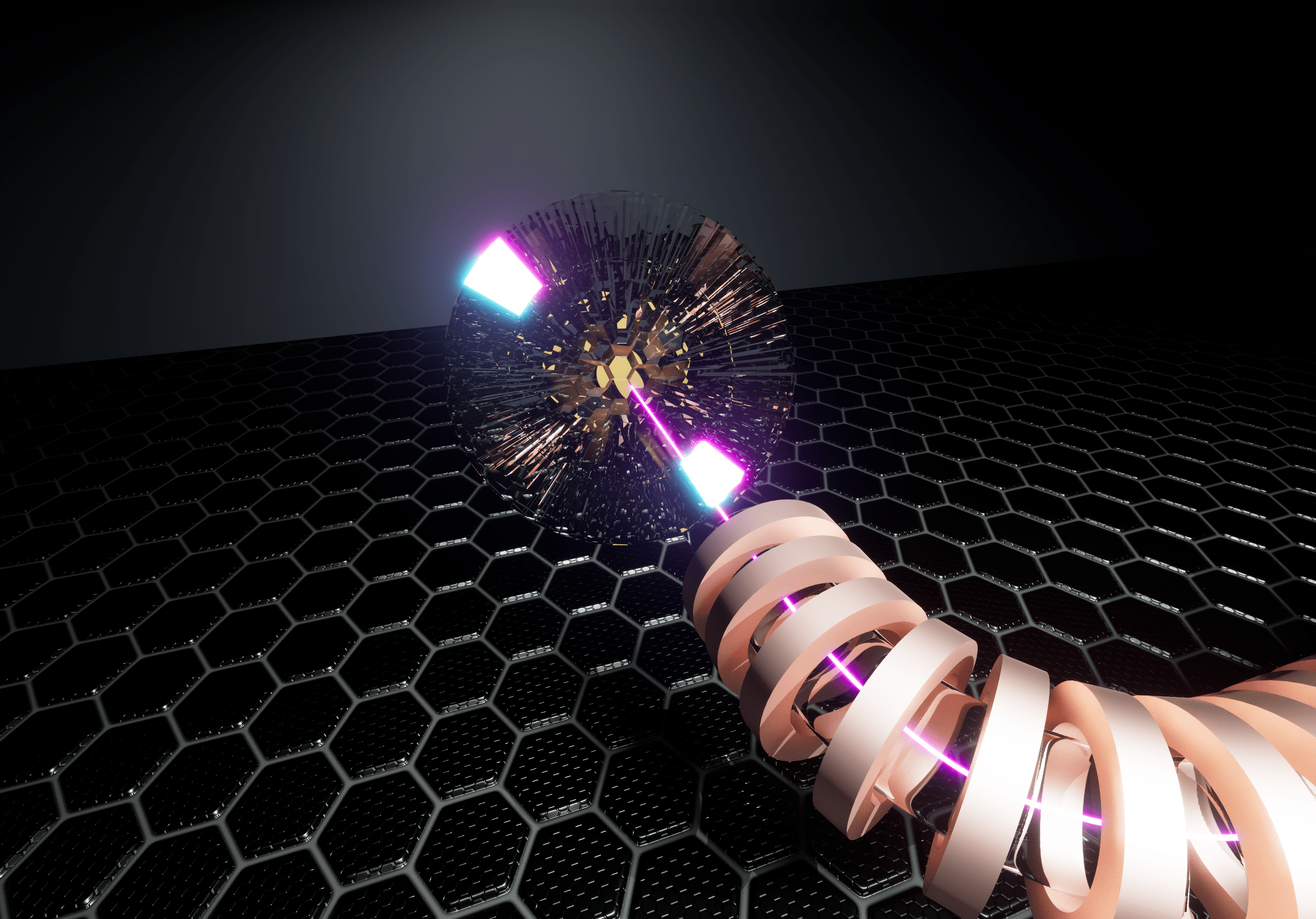}}
    \caption{Schematic view of the ECAL geometry.}
    \label{fig:cal}
\end{figure}

The Polygon Mesh Processing Library (PMP)~\cite{Sieger2019} is used to implement the above algorithm, which generates the ECAL geometry in the MACE offline software.
A Class I GP$(8,0)$ Goldberg polyhedron mesh inscribed in the unit sphere is produced, consisting of a total of 642 faces, including 12 regular pentagons and 630 irregular hexagons, which can be categorized into 9 types.
By scaling the obtained mesh using the ECAL inner radius $\boldsymbol{r}$ and the crystal length $\boldsymbol{l}$, a parameterized ECAL geometry which can be easily adjusted is established in the MACE offline software via \texttt{G4TessellatedSolid} in \textsc{Geant4}.
Modules created from the different types of faces are identified as Type-PEN and Type-HEX (01--09).
\cref{fig:cal} displays the ECAL geometry rendered using Unreal Engine \cite{Unreal}.

\begin{table}[!t]
    \centering
    \caption{Properties of several inorganic scintillation crystals~\cite{ParticleDataGroup:2024cfk,Hu:2022rbc}}.
    \label{tab:crystal}
    \renewcommand{\arraystretch}{0.6}
    \setlength{\tabcolsep}{5pt}
    \begin{threeparttable}[t]
        \begin{tabular}{lccccccccc}
            \hline \hline
            \specialrule{0em}{2pt}{2pt}
            \textbf{Parameter:}                                                         &
            $\rho$                                                                      &
            $X_0$                                                                       &
            $R_M$                                                                       &
            $dE/dx$                                                                     &
            $\tau_{decay}$                                                              &
            $\lambda_{max}$                                                             &
            $n$                                                                         &
            \multirow{2}{*}{\begin{tabular}[c]{@{}c@{}}Relative \\ output\end{tabular}} &
            \multirow{2}{*}{\begin{tabular}[c]{@{}c@{}}Hygro-\\ scopic\end{tabular}}                                                                                                        \\
            \specialrule{0em}{5pt}{5pt}
            \textbf{Units}                                                              & g$/\text{cm}^3$ & cm   & cm   & MeV$/$cm & ns            & nm  &      &                  &        \\ \hline
            \specialrule{0em}{3pt}{3pt}
            NaI(Tl)                                                                     & 3.67            & 2.59 & 4.13 & 4.8      & 245           & 410 & 1.85 & 100              & yes    \\
            BGO                                                                         & 7.13            & 1.12 & 2.23 & 9.0      & 300           & 480 & 2.15 & 21               & no     \\
            CsI(Tl)                                                                     & 4.51            & 1.86 & 3.57 & 5.6      & 1220          & 550 & 1.79 & 165              & slight \\
            CsI(pure)                                                                   & 4.51            & 1.86 & 3.57 & 5.6      & 30\tnote{$s$} & 420 & 1.95 & 3.6\tnote{$s$}   & slight \\
                                                                                        &                 &      &      &          & 6\tnote{$f$}  &     &      & 1.1\tnote{$f$}   &        \\
            PbWO$_4$                                                                    & 8.30            & 0.89 & 2.00 & 10.1     & 30\tnote{$s$} & 425 & 2.20 & 0.3\tnote{$s$}   & no     \\
                                                                                        &                 &      &      &          & 10\tnote{$f$} & 420 &      & 0.077\tnote{$f$} &        \\
            LYSO(Ce)                                                                    & 7.40            & 1.14 & 2.07 & 9.6      & 40            & 420 & 1.82 & 85               & no     \\
            CAGG(Ce)                                                                    & 6.50            & 1.63 & 2.20 & 9.0      & 53            & 540 & 1.92 & 98               & no     \\
            LaBr$_{3}$(Ce)                                                              & 5.29            & 1.88 & 2.85 & 6.9      & 20            & 356 & 1.90 & 180              & yes    \\                   \specialrule{0em}{2pt}{2pt} \hline \hline
        \end{tabular}%
        \begin{tablenotes}
            \footnotesize
            \item $^f$ fast component, $^s$ slow component
        \end{tablenotes}
    \end{threeparttable}
\end{table}

Calorimeters with inorganic scintillation crystals generally provide excellent energy resolution and thus have been widely used in particle physics experiments for decades.
Commonly used scintillators in calorimeters include NaI(Tl), pure CsI, CsI(Tl), BGO, PbWO$_4$, and others \cite{Song2023}.
\cref{tab:crystal} summarizes the properties of these inorganic scintillators.
The scintillation crystal selection for the MACE ECAL is based on the energy region of interest less than 1~MeV.
NaI(Tl) was widely used in the past decades due to its low cost and high scintillation light output.
Nonetheless, it is hygroscopic and has a relatively low density, making it less suitable for large-scale particle physics experiments at present. The problem of hygroscopy is identical for LaBr$_3$(Ce). Moreover, considering the overall budget of this experiment, ultrafast and bright crystals such as LYSO(Ce), GAGG(Ce), or LaBr$_3$(Ce) are not currently included but may be considered for potential future upgrades.
The low light output of PbWO$_4$ leads to an insufficient energy resolution in the low-energy regime. The same is true for pure CsI.
In the MACS experiment, the energy resolution of the pure CsI calorimeter for single-$\gamma$ event and double-$\gamma$ events was 69\% and 52\%, respectively~\cite{Willmann1995,Willmann1999}.
Doping thallium (Tl) in pure CsI can significantly increase its light yield, making it one of the brightest scintillators known.
The objective of MACE ECAL is to achieve excellent energy resolution for $\gamma$-rays produced in $e^+e^-$ annihilation. This requires optimizing the light yield of the scintillator, which is a key factor in precise energy measurements.
Consequently, CsI(Tl) is chosen as the reference scintillator in this ECAL design.

The photosensor is also a crucial component in the calorimeter system.
Its technical specifications (e.g., sensitive area, photon detection efficiency) determine the efficiency with which optical photons are converted into electrical signals that can ultimately be recorded.
Commonly used photosensors in current particle and nuclear physics experiments include Photomultiplier Tubes (PMTs) and Silicon Photomultipliers (SiPMs)~\cite{Simon2018}.
PMTs are relatively more cost-effective than SiPMs for the same sensitive area but are constrained by the low quantum efficiency (QE), high operating voltage, and shielding requirements in magnetic fields.
On the other hand, SiPMs offer excellent photon detection efficiency (PDE), a compact size, and tolerance to magnetic fields. However, they come with higher costs, dark current due to semiconductor thermal noise, and crosstalk between adjacent pixels~\cite{Gundacker2020}.
Currently, PMTs are considered the benchmark photosensor in this paper.
A preliminary test of the SiPM performance is also discussed in~\cref{sec:discussion}.

\subsection{Optimization of Geometry Parameters}
The ECAL geometry is generated by expanding the Goldberg polyhedron mesh radially, with the ECAL inner radius and crystal length determining the module size.
In a scintillator-based detection system, the energy resolution is primarily determined by the scintillation light yield and its collection efficiency.
Additionally, the geometric dimensions of each module are crucial for light collection efficiency~\cite{Lecoq2017}, considering factors such as the $\gamma$-ray energy deposition rate and the propagation of optical photons within the crystal, which are ultimately collected by the photosensor.
To enhance the sensitivity of ECAL, the energy resolution should be optimized by scanning geometry parameters, including the crystal length and the ECAL inner radius.

\begin{figure}[!b]
    \centering
    \subfloat[Results of photon counts.]{\includegraphics[width=.33\linewidth]{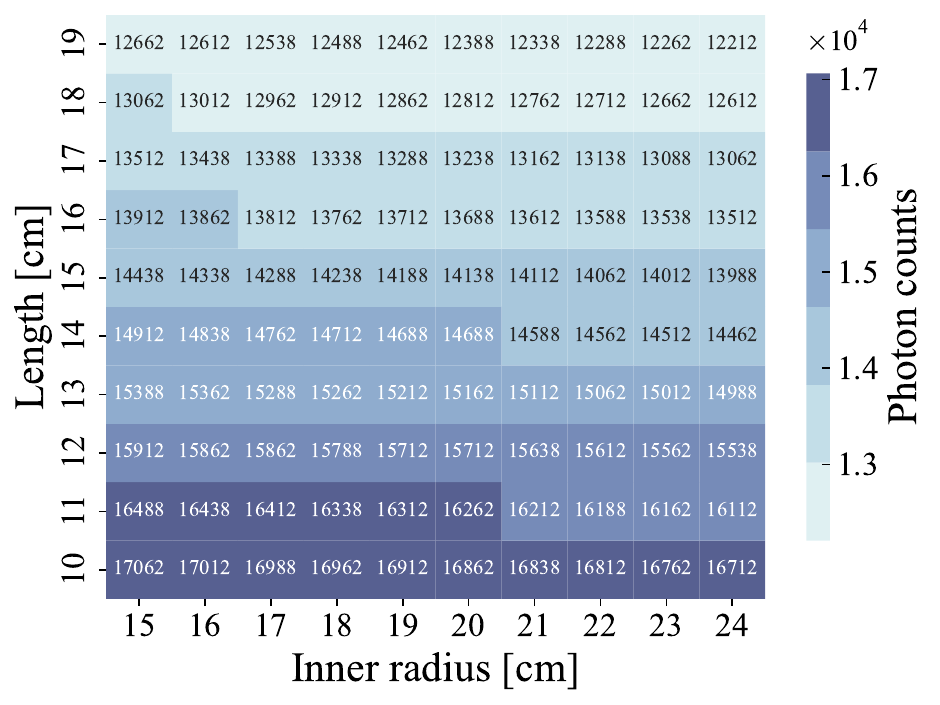}}
    \subfloat[Results of detection efficiency.]{\includegraphics[width=.33\linewidth]{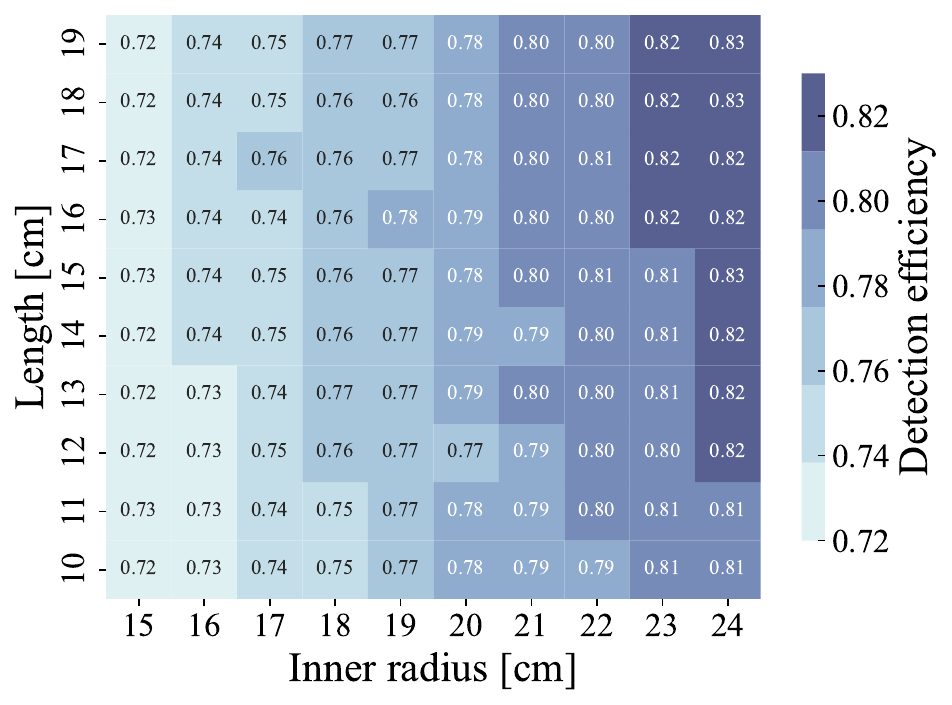}}
    \subfloat[Results of F.o.M.]{\includegraphics[width=.33\linewidth]{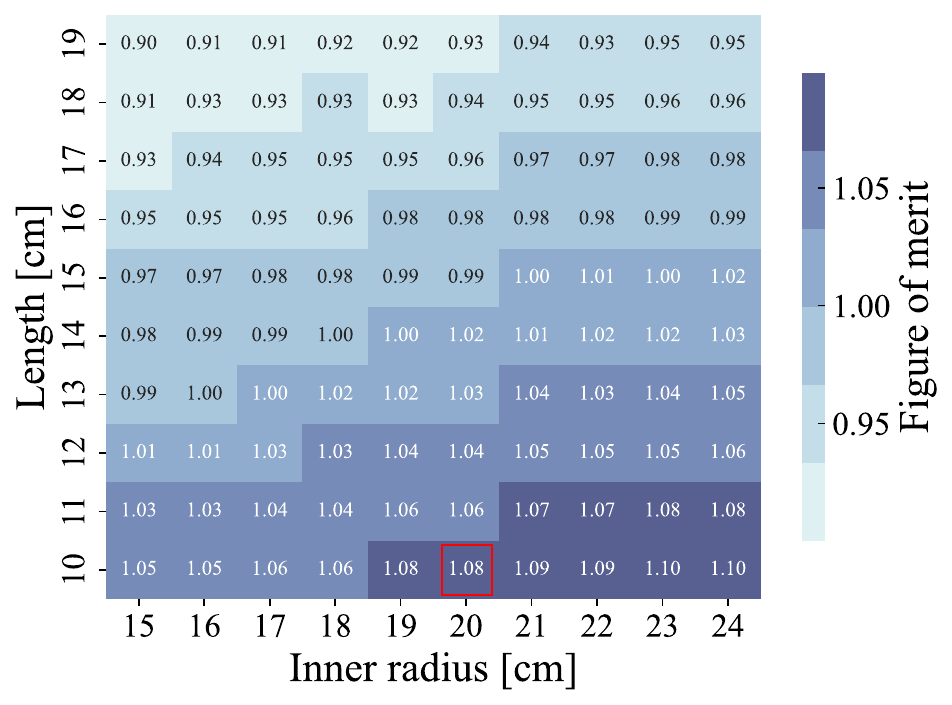}}
    \caption{Scanning results of collected photon counts at maximum and detection efficiency of Type-HEX06 module with different geometry parameters including crystal length and ECAL inner radius. A deeper color suggests a better performance in all figures.}
    \label{fig:ECAL-scan}
\end{figure}

A detailed Monte-Carlo simulation of the module Type-HEX06, which is characterized by both the largest number and size, is performed using MACE offline software based on \textsc{Geant4} v11.2.2.
The module mentioned above and below is defined as a combination of a scintillation crystal and a photosensor assembled together.
The physics processes are modeled with the reference physics list \texttt{QBBC\_EMZ}, \texttt{G4OpticalPhysics}, and the \texttt{Unified} surface model.
Additionally, the optical properties of Teflon-wrapped CsI(Tl) crystals (such as light yield, emission spectrum, decay time, etc.) and the PDE of the photosensor are considered based on manufacturer data sheets.
Detailed configurations of the simulation are presented in~\cref{tab:properties}.
Note that to better simulate the energy resolution of CsI(Tl), a Fano Factor $F_\text{N}$ is introduced~\cite{Bora2016,Yanagida:2022gws}, and its squared value is assigned to the \texttt{RESOLUTIONSCALE} parameter in \textsc{Geant4}.
Monoenergetic 511~keV $\gamma$-rays from a point-biased source are directed onto the center of a crystal optically coupled with a photosensor, which is registered as a Sensitive Detector (SD).
The area of SD is adjusted to fully cover the outer surface of a Type-HEX06 crystal to estimate the intrinsic energy resolution of the crystal.
The crystal length varies from 10~cm to 19~cm, and the inner radius of ECAL varies from 15 cm to 24~cm.
Peak finding is performed using \texttt{TSpectrum} within the ROOT framework~\cite{Brun1997} to record the maximum count of scintillation photons spectra collected by the SD.
Meanwhile, the detection efficiency is given by the ratio of the number of events with energy greater than 500~keV to the total number of incident particles. The results of scanning each geometry parameter combination are detailed in \cref{fig:ECAL-scan}.

\begin{table}
    \centering
    \caption{Specification of material properties.}
    \label{tab:properties}
    \renewcommand{\arraystretch}{0.8}
    \begin{tabular}{lccccc}
        \hline\hline
                    & Material & Density                  & Refractive index   & Absorption length  & Reflectivity       \\
                    &          & $\rho$ (g$/\text{cm}^3$) & $n$                & $\lambda$ (cm)     & $R$                \\
        \hline
        Crystal     & CsI(Tl)  & 4.51                     & 1.79               & 37                 & -                  \\
        \hline
        Reflector   & Teflon   & 2.25                     & 1.35               & -                  & 0.98               \\
        \hline
        Coupler     & EJ-550   & 0.97                     & 1.46               & 100                & \multirow{3}{*}{-} \\
        \cline{1-5}
        PMT window  & Glass    & 2.40                     & 1.49               & \multirow{4}{*}{-} &                    \\
        \cline{1-4}
        SiPM window & Epoxy    & 1.18                     & 1.55               &                    &                    \\
        \cline{1-4}\cline{6-6}
        PMT cathode & Bialkali & 2.00                     & \multirow{2}{*}{-} &                    & \multirow{2}{*}{0} \\
        \cline{1-3}
        SiPM sensor & Silicon  & 2.33                     &                    &                    &                    \\
        \hline\hline
    \end{tabular}
\end{table}

It should be noted that there is a trade-off between energy resolution and detection efficiency.
The result demonstrates that increasing of the crystal volume causes a decrease in light collection efficiency and a deterioration in energy resolution, as investigated by Refs. \cite{Huber1999,Mitra2019}.
It is reasonable to assume that, as optical photons travel a longer distance, some of them will be absorbed by the material directly or after multiple reflections, resulting in reduced light collection efficiency.
On the other hand, the crystal length has a limited effect on the detection efficiency because the longitudinal leakage of 511 keV $\gamma$-rays is negligible.
Incident particles are more likely to escape laterally due to Compton scattering; thus, detection efficiency increases with the ECAL inner radius, which affects the module width.
To balance the energy resolution and the detection efficiency, a Figure-of-Merit (F.o.M) function is constructed, following the function used in Ref.~\cite{Papa:2023voa}:
\begin{equation}
    \text{F.o.M}=\sqrt{\frac{N_{\text{pho}}}{\langle N_{\text{pho}}\rangle}\cdot\frac{\eta}{\langle\eta\rangle}}~,
\end{equation}
where $N_{\text{pho}}$ is the maximum count of scintillation photons spectra, $\eta$ is the detection efficiency. These two quantities are normalized by their mean values.
The geometric parameters are ultimately chosen to be a crystal length of 10 cm and an ECAL inner radius of 20 cm based on consideration of both F.o.M and spatial constraints.

\begin{figure}[!t]
    \centering
    \includegraphics[width=.7\linewidth]{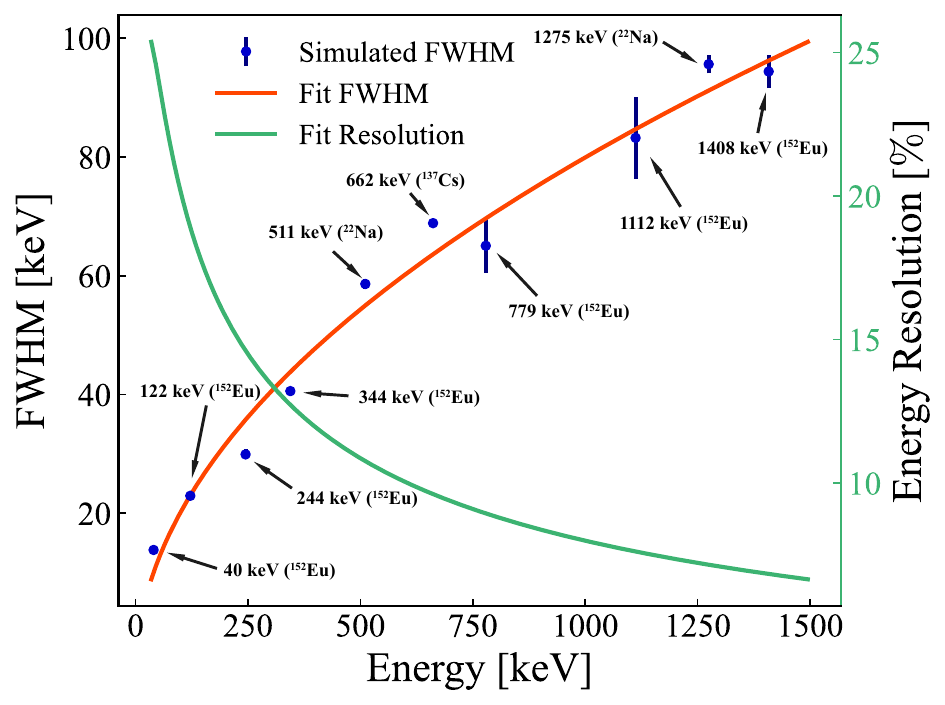}
    \caption{Calibration by \textsc{Geant4} simulation of Type-HEX06 module. The red line is the FWHM curve fit by simulation data (\textit{blue dots with error bars}), while the green line is the fit energy resolution. The errors are introduced by fitting the full-energy peak.}
    \label{fig:fwhm_emc}
\end{figure}

In the context of the MACE experiment, the signal positrons from $\bar{\text{M}}$ decay are transported through a solenoid into the region of the positron detection system.
Some modules must be removed to ensure that the incoming signal positrons are not obstructed.
Specifically, the beam windows are set with an upstream radius of 50 mm and a downstream radius of 5~mm.
Consequently, 19 modules at the beam entrance and 1 module at the exit are removed, resulting in a total of 622 modules and achieving 97\% coverage of the $4\pi$ solid angle.
For an estimation of energy resolution, the class \texttt{G4RadioactiveDecayPhysics} is used to simulate the calibration of the Type-HEX06 module with radioactive sources, including $^{22}$Na, $^{137}$Cs, and $^{152}$Eu.
Instead of the fully-covered SD in the previous simulation, the crystal is coupled with a 1.5-inch diameter PMT volume.
An isotropic point source, wrapped in 1 mm thick plastic, is placed 5~cm away from the module.
The energy dependence of FWHM is fit using~\cref{eq:fwhmFormula}:
\begin{equation} \label{eq:fwhmFormula}
    \text{FWHM}=a+b\sqrt{E_0}~,
\end{equation}
where the best-fit value of $a$, $b$ are $-7.4707$ and $2.7638$, respectively. The fit FWHM and energy resolution results of each full-energy peak are plotted in~\cref{fig:fwhm_emc}.
For the Type-HEX06 modules, the energy resolution is 10.8\% at 511 keV, which is 6--7 times better than the resolution achieved by the calorimeter used in the MACS experiment~\cite{Willmann1999}.

\subsection{Signal and Backgrounds}
\subsubsection{M-to-$\bar{\text{M}}$ conversion signal}
In the M-to-$\bar{\text{M}}$ conversion process, $\bar{\text{M}}$ decays into a Michel electron and an atomic-shell positron.
The positron is then accelerated electrostatically and transported through the solenoid into the ECAL chamber, where it is detected by the MCP.
Due to the low kinetic energy of the positron, it will annihilate in MCP, producing a pair of back-to-back $\gamma$-rays.
The expected signal for the positron in M-to-$\bar{\text{M}}$ conversion event will be identified through a coincidence of MCP and ECAL responses.
This section details the coincidence studies conducted using the MACE offline software, which provides insights into event reconstruction and sensitivity evaluation.

A beam of \(10^8\) \(\mu^+\) events with a momentum of 26.3~MeV$/c$ entering the MACE detector has been simulated.
The \(\mu^+\) forms M in the production target with a certain probability, which then completely converts into $\bar{\text{M}}$ which subsequently decays.
Event sampling is performed according to the detection efficiency curve of MCP, and positron events with multiple hits on MCP are filtered out.
After a positron annihilates, the produced \(\gamma\)-rays either interact with the sensitive volume of ECAL or escape the chamber.
The coincidence between MCP and ECAL responses ensures that the annihilation signals originate from the same M-to-$\bar{\text{M}}$ conversion event.
Following the coincidence, a clustering algorithm is applied to search seed modules and their adjacent modules.
The energy and direction information of the hits is reconstructed, and only two simultaneous responses that satisfy certain criteria will be selected.
The event selection criteria can be summarized as follows:
\begin{itemize}
    \item \textbf{Trigger threshold}: $E_{\text{Thr}}>2\sigma$;
    \item \textbf{Energy difference cut}: $\Delta E_{\text{Recon}}<3\sigma$;
    \item \textbf{Angle cut}: $\theta_{\text{Recon}}>170^{\circ}$.
\end{itemize}
These criteria stand for trigger threshold, energy difference, and angle of two reconstructed tracks, respectively.
Finally, the ECAL response spectra are smeared using the FWHM curve described by~\cref{eq:fwhmFormula}. Events that fall within the \(3\sigma\) interval of the full-energy peak are identified as signal events.

\begin{figure}[htb]
    \centering
    \includegraphics[width=0.7\linewidth]{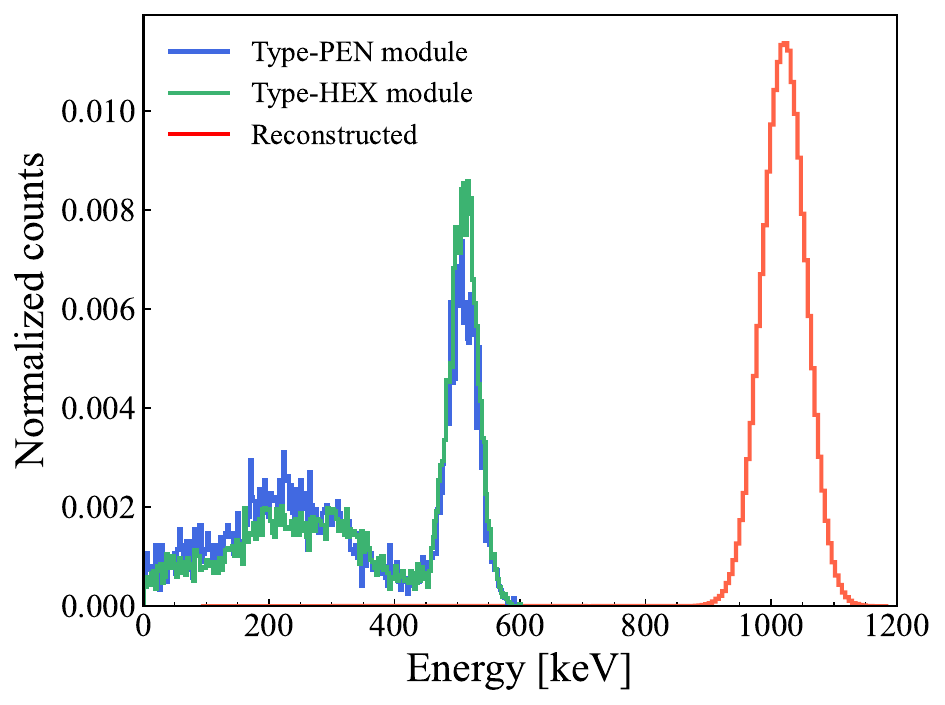}
    \caption{Spectra of a single Type-PEN module (\textit{blue line}), a single Type-HEX module (\textit{green line}) and reconstructed annihilation signals (\textit{red line}) of M-to-$\bar{\text{M}}$ conversion by simulation.}
    \label{fig:signal}
\end{figure}

\cref{fig:signal} summarizes the simulated spectra of the positron annihilation signal.
In the original single-module $\gamma$ spectrum, the 511~keV full-energy peak accounts for only 47.2\% of the total counts.
This is attributed to some $\gamma$-rays scattering to adjacent modules after depositing part of their energy, which forms a Compton continuum.
As a result, the signal efficiency will be significantly reduced.
However, by clustering the energy responses from each event, the total energy of the scattered photons can be reconstructed.
As can be seen from~\cref{fig:signal}, the energy has been reconstructed effectively.
The expected signal is thus represented by the \(3\sigma\) interval of 1.022~MeV.

\begin{table}[!t]
    \centering
    \caption{Summary of factors contributing to the signal efficiency for double $\gamma$ events of the ECAL.}
    \label{tab:efficiency}
    \begin{tabular}{lcc}
        \hline\hline
        \multicolumn{2}{l}{\textbf{Items}}                                                                       & \textbf{Efficiency}                      \\
        \hline
        \multicolumn{2}{l}{Geometric efficiency of single $\gamma$ event $\varepsilon^{1\gamma}_{\text{Geom}}$}  & 95.2\%                                   \\
        \multicolumn{2}{l}{Geometric efficiency of double $\gamma$ events $\varepsilon^{2\gamma}_{\text{Geom}}$} & 93.6\%                                   \\
        \hline
        \multicolumn{2}{l}{Energy over threshold $\varepsilon_{\text{Trig}}$ ($E_{\text{Thr}}>2\sigma$)}         & 98.7\%                                   \\
        \hline
        \multirow{2}{*}{Reconstruction efficiency $\varepsilon_{\text{Recon}}$}                                  & $\Delta E_\text{Recon}<3\sigma$ & 90.8\% \\
        \cline{2-3}
                                                                                                                 & $\theta_\text{Recon}>170^\circ$ & 86.0\% \\
        \hline
        \multicolumn{2}{l}{Total signal efficiency}                                                              & 78.3\%                                   \\
        \hline\hline
    \end{tabular}
\end{table}

The result of the simulation indicates a 78.3\% signal efficiency for double $\gamma$ events in total.
The factors contributing to the total signal efficiency are summarized in~\cref{tab:efficiency}, where the standard deviation in the full-energy peak $\sigma$ is determined by the energy resolution.
These results suggest that the efficiency performance of the ECAL surpasses that of the MACS calorimeter, which achieves an $\varepsilon^{1\gamma}$ of 79\%.

\subsubsection{Beam-related backgrounds}
The muon beam, produced by stopping \(\pi^+\) on the surface of a target, often includes positrons or other particles with similar momentum \cite{Xu2024}.
These background particles can enter the ECAL chamber in the same manner as the signal positrons, potentially causing accidental coincidences between the MCP and the ECAL.
Therefore, it is crucial to estimate their potential contribution.

A parallel simulation involving \(10^{11}\) \(e^+\) events was conducted for comparison with the signal.
No coincidence events were observed in this simulation.
According to the technical specifications of the Experimental Muon Source (EMuS) proposed at CSNS~\cite{Hong2024}, the optimal ratio of positrons to muons is expected to be 0.1\%~\cite{Zhou2022}.
The beam positron background level is estimated based on this data.
Assuming a \(\mu^+\) beam flux of \(10^8/\)s, and a coincidence probability of \(10^{-4}\) with the magnetic spectrometer, the upper limit of the background rates is calculated to be $0.007/(10^8~\mu^+/\text{s}\cdot1~\text{yr})$ at 90\%~C.L.

\subsubsection{Cosmic-Ray Muon backgrounds}
Cosmic-ray muons are another source of background in ECAL.
The EcoMug generator~\cite{Pagano2021} is integrated into the MACE offline software to simulate cosmic-ray muon events on various types of surfaces (plane, cylinder, and half-sphere) with accurate angle and momentum distributions.
To evaluate the cosmic-ray muon background level in a potential future environment for the MACE detector, a hypothetical 18~m deep tunnel has been modeled in \textsc{Geant4}.
A concrete volume, represented by \texttt{G4Box} with dimensions $50 \times 50 \times 18~\text{m}^3$, is created, with an air-filled cylindrical region, represented by \texttt{G4Tubs}, subtracted from its bottom to form a tunnel with a 1 m radius and a 10 m length.
ECAL is positioned at the center of the tunnel, surrounded by a copper solenoid and an iron shield.
Cosmic-ray muons are emitted from a $50 \times 50~\text{m}^2$ plane source at sea level.
With a rate of 129 Hz, cosmic-ray muon events equivalent to 1 hour of exposure are generated.

The simulation results indicate that cosmic-ray muons still contribute to the background in the ECAL even after passing through the concrete and shielding. This includes some events where $e^+e^-$ annihilation produces 511~keV $\gamma$ rays.
After applying coincidence selection, no events remain, resulting in a background rate of less than $2/$yr when considering coincidences with the Magnetic Spectrometer.
Typically, a veto system can be installed outside ECAL to further reject cosmic-ray muon background~\cite{Abramishvili2020,Shah2024}.

\subsubsection{Radioactive contamination background}
It is worth mentioning that studies have reported the radioactive contamination of scintillation materials (e.g., $^{137}$Cs in CsI), which may greatly affect low-counting experiments. In the context of the MACE PDS, if the MCP is triggered by its dark counts while the radionuclides within the ECAL materials decay simultaneously, a false coincidence may occur. Based on the provided data, a single CsI(Tl) module in the ECAL will have a 662~keV $\gamma$-ray background counting rate up to 0.07~Hz~\cite{Danevich:2018wmq}. A preliminary estimation of the accidental coincidence rate can be conducted by
\begin{equation}
    \begin{gathered}
        n=2\tau n_{\text{MCP}}n_{\text{ECAL}}~,\\
        n_{\text{ECAL}}=2\tau n_{\text{mod1}}n_{\text{mod2}}\times N_{\text{mod}}/2~,
    \end{gathered}
\end{equation}
where $\tau$ is the coincidence resolving time which is set to be CsI(Tl) decay time, $n_{\text{MCP}}$ is the dark count rate of the MCP (commonly $\sim1$ kHz), $n_{\text{ECAL}}$ is the accidental coincidence of two modules caused by the natural background detected by the ECAL and $N_{\text{mod}}$ is the total module numbers. Eventually, the rate is calculated to be $0.19/$yr. This background, however, falls outside the $3\sigma$ interval of the signal of interest. If the event selection criteria are applied, the background caused by dark count and natural radioactivity can be neglected.

\section{Validation with a Benchmark Detector} \label{sec:3}
\subsection{Experiment Setups and Results}
\begin{figure}[!t]
    \centering
    \subfloat[CsI(Tl) crystal.]{\label{subfig:crystal}\includegraphics[width=0.48\linewidth]{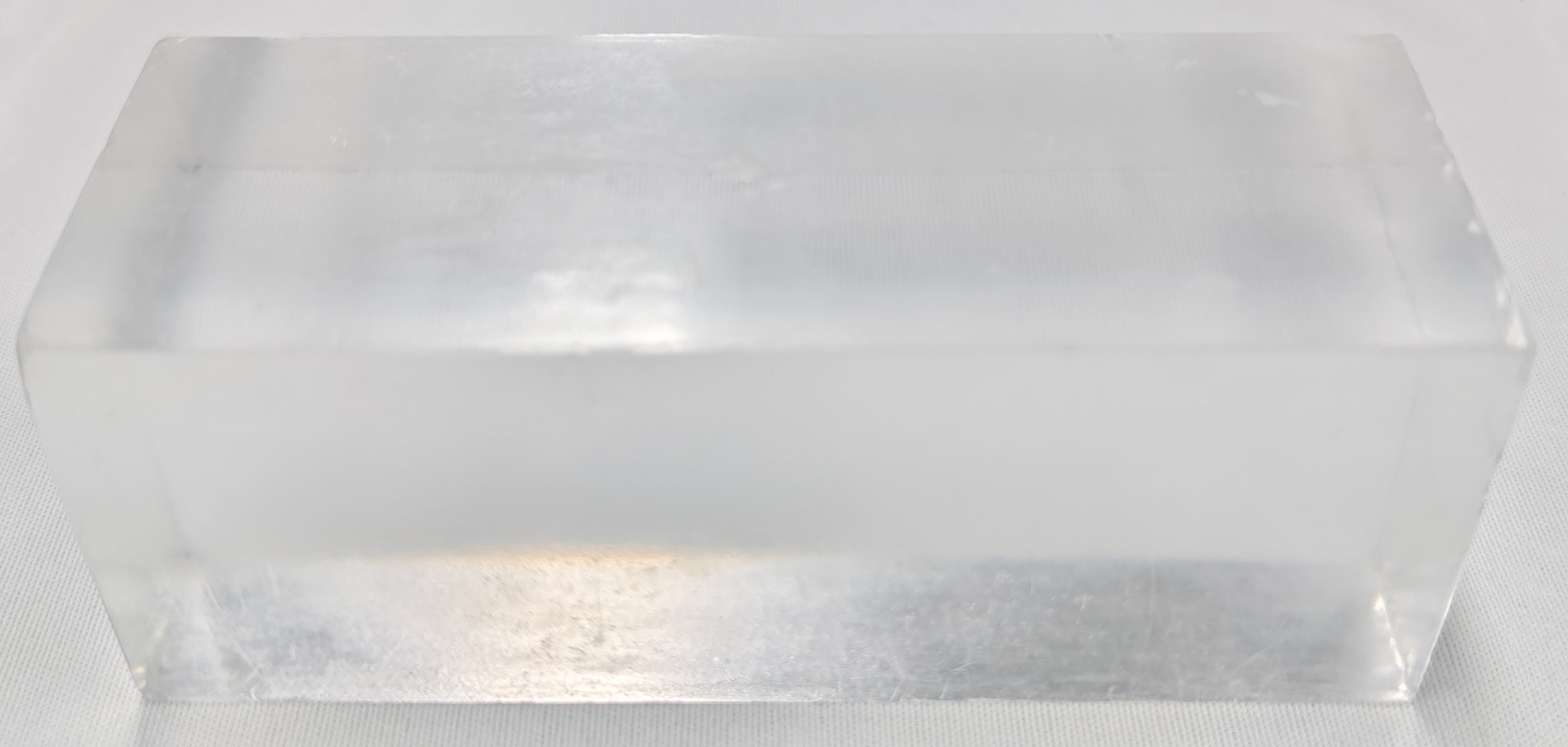}}
    \subfloat[Benchmark detector.]{\label{subfig:benchmark} \includegraphics[width=0.48\linewidth]{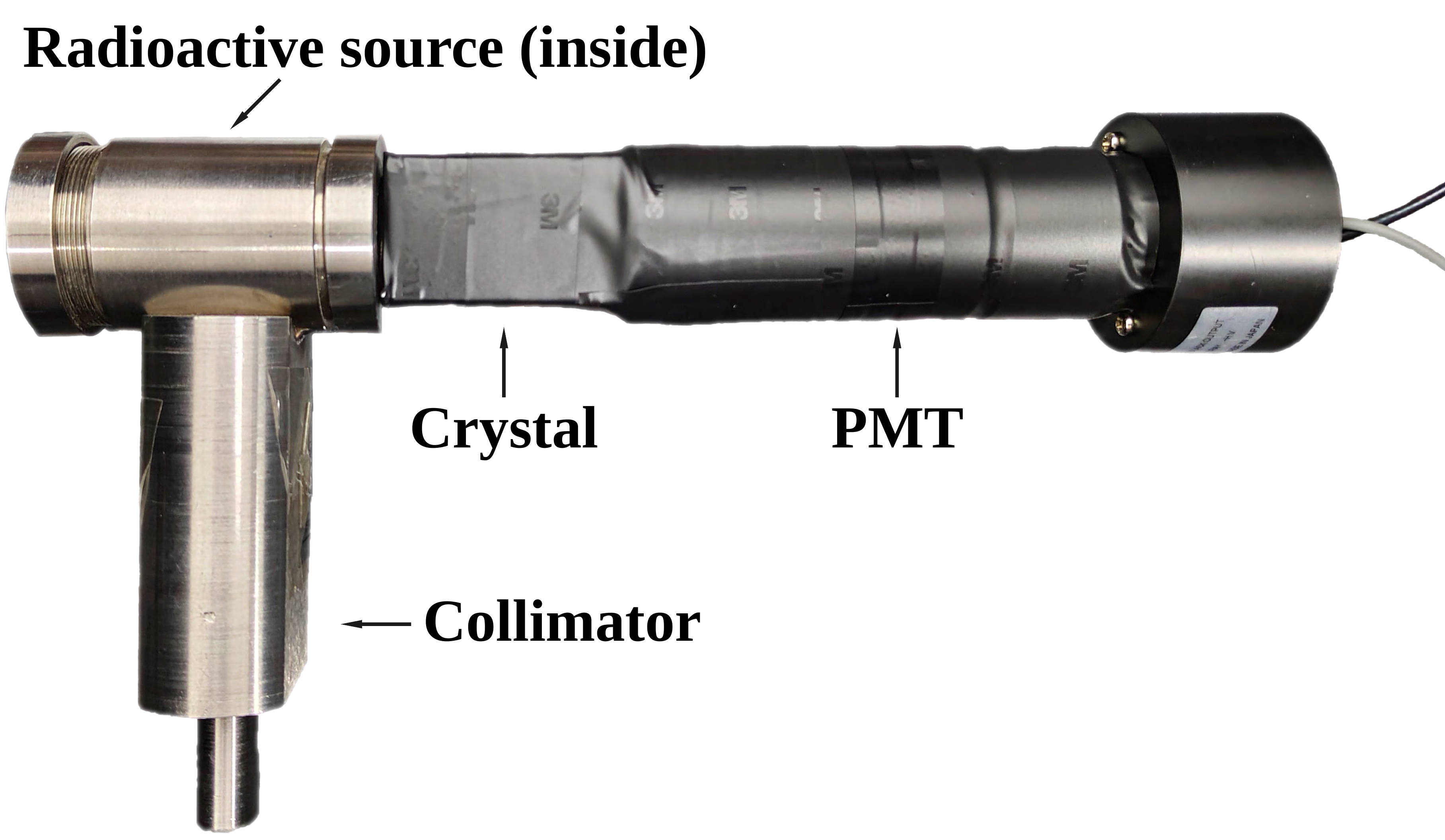}}
    \hfil
    \subfloat[Lecroy HDO4054A displaying a typical waveform (pulse width of $\sim2~\mu$s) of CsI(Tl).]{\label{subfig:oscilloscope}\includegraphics[width=0.48\linewidth]{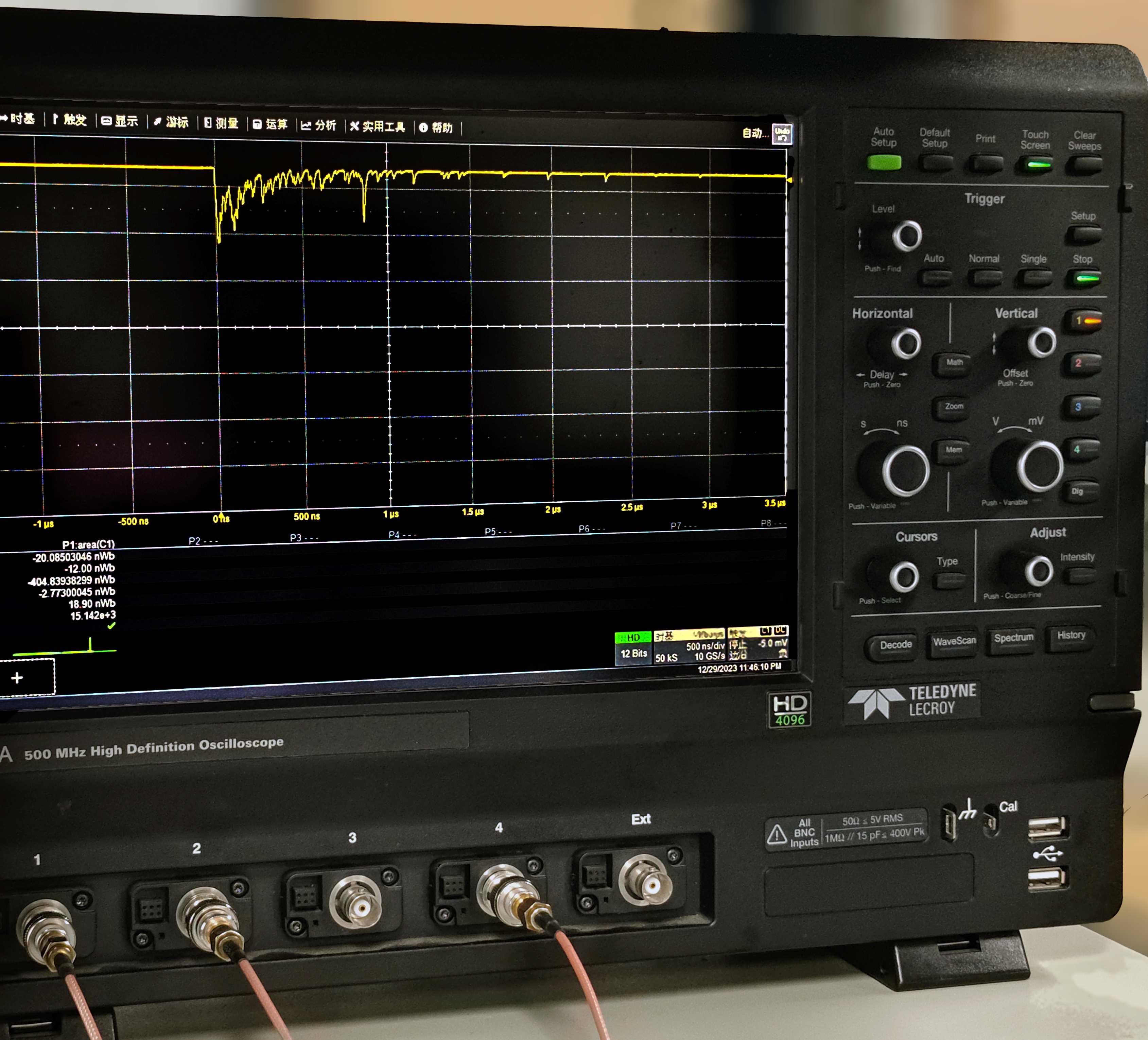}}
    \subfloat[CAEN V1730SB digitizer and the VMEbus.]{\label{subfig:digitizer}\includegraphics[width=0.48\linewidth]{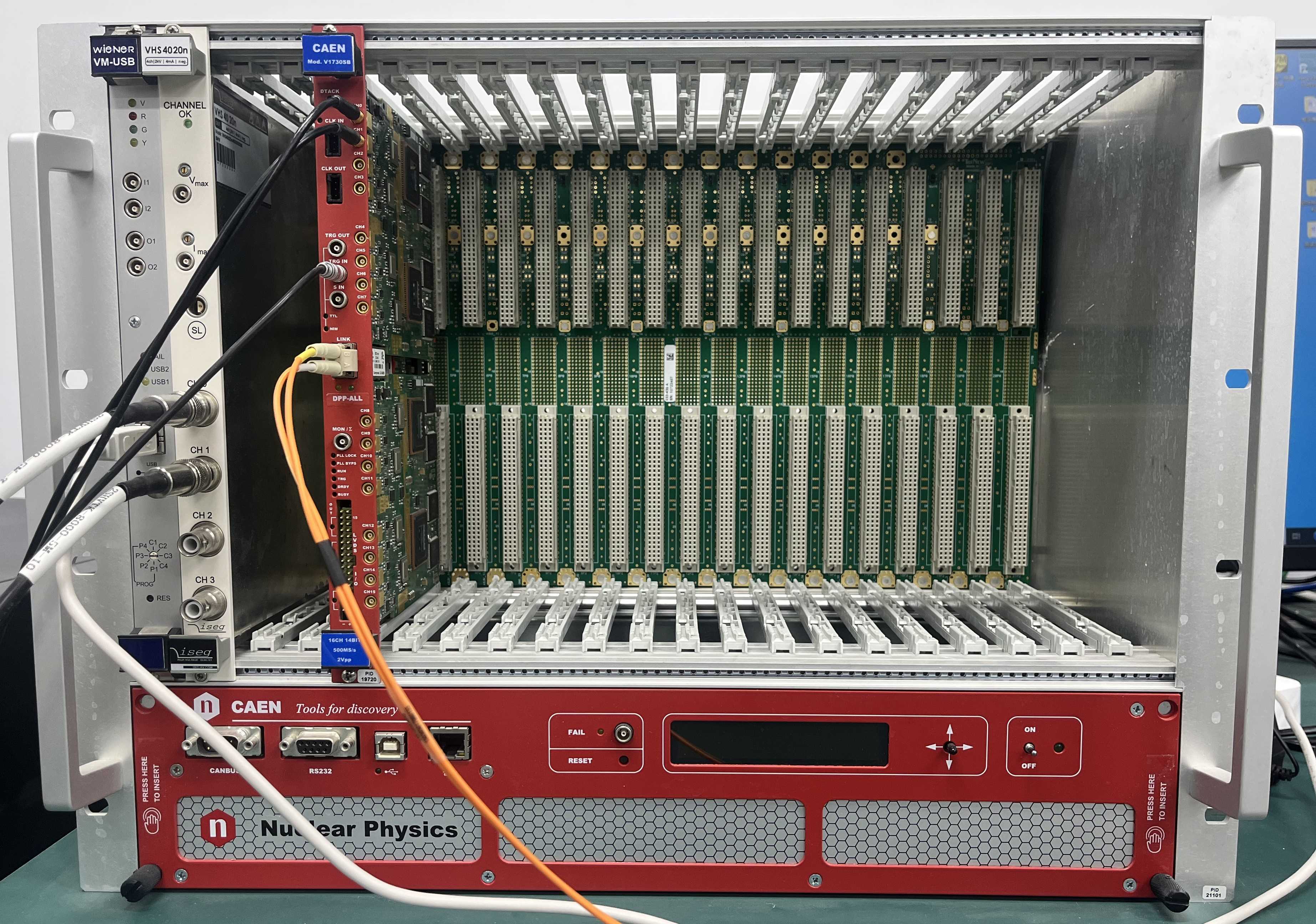}}
    \caption{Photographs of the experiment setups.}
    \label{fig:benchmarkPhoto}
\end{figure}

\begin{figure}[htbp]
    \centering
    \includegraphics[width=0.7\linewidth]{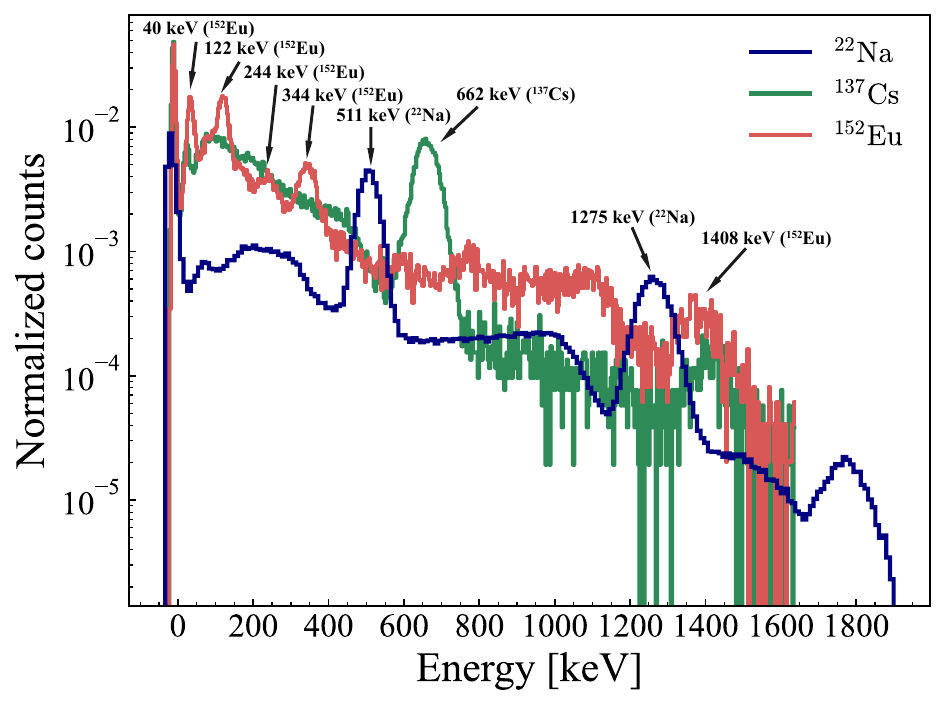}
    \caption{Measured spectra of $^{137}\text{Cs}$ (\textit{green line}), $^{152}\text{Eu}$ (\textit{red line}) at SYSU and $^{22}\text{Na}$ (\textit{blue line}) at IMP.}
    \label{fig:spec_exp}
\end{figure}

To characterize the parameters of the CsI(Tl) scintillator, a benchmark detector was constructed and calibrated using various radioactive sources at Sun Yat-sen University (SYSU).
It consists of a $3 \times 3 \times 8~\text{cm}^3$ CsI(Tl) crystal and a 1.5-inch PMT (HAMAMATSU CR284 with E2183-500 socket assembly).
The crystal is wrapped with ESR (glued) and coupled with PMT using optical grease, finally coated with 3M tape (\cref{subfig:benchmark}).
The readout system uses a Lecroy HDO4054A oscilloscope (12~bit, 10~GS/s, see \cref{subfig:oscilloscope}) for signal triggering, integration, and digitization.
The benchmark detector is calibrated using $^{137}\text{Cs}$ ($7.81\times10^3$~Bq) and $^{152}\text{Eu}$ ($3.55\times10^4$~Bq) radioactive sources.
Radioactive sources were positioned 5~cm away from the detector, and data acquisition was performed for 1800~s.
When the PMT pulse signal exceeds the threshold, the oscilloscope triggers, integrates the waveform over a 2~$\mu$s time window, and records the charge value.
The calibration used the 662 keV peak from $^{137}\text{Cs}$ and the 40 keV, 122 keV, 244 keV, 344 keV, and 1408 keV peaks from $^{152}\text{Eu}$, as shown in \cref{fig:spec_exp}. The spectra display the full-energy peaks of both isotopes.
The energy resolution at 511~keV is calculated to be 11.5\% by fitting the FWHM curve.

A more precise experiment was conducted to investigate the energy resolution at the Institute of Modern Physics (IMP), Chinese Academy of Sciences in Lanzhou, China.
The absolute gain of the CR284 PMT was accurately calibrated in advance using single photoelectron (SPE) response measurements.
A waveform generator sends pulses with a voltage of 2.65~V, a frequency of 10~kHz, and a width of 5~ns to an LED to produce weak light emission.
A 120~s measurement scans the high voltage ranging from 850~V to 1250~V.
The PMT response waveforms are read out and recorded in ROOT files using the CAEN V1730SB digitizer (14~bit, 500~Ms/s, see \cref{subfig:digitizer}), which is externally triggered by the waveform generator.
After offline pulse processing, the SPE peaks are fit to determine their mean charge value which is then converted to the absolute gain.

\begin{table}[htbp]
    \centering
    \caption{Energy resolution (FWHM) performance at 511 keV of the benchmark detector with various reflectors.}
    \label{tab:reflector}
    \begin{tabular*}{0.7\linewidth}{@{\extracolsep{\fill}}cc}
        \hline \hline
        \textbf{Reflector} & \textbf{Energy resolution} \\ \hline
        Teflon                     & 10.3\%                                        \\
        ESR (non-glued)              & 11.1\%                                        \\
        ESR (glued)              & 11.5\%                                        \\
        Tyvek                      & 13.1\%                                        \\
        \hline \hline
    \end{tabular*}
\end{table}

Subsequently, the energy resolution performance with various reflectors is tested. The CsI(Tl) crystal is wrapped with materials including Teflon, ESR (non-glued), and Tyvek. Other experimental setups remained identical to \cref{subfig:benchmark}, except for the use of $^{22}\text{Na}$ ($1.48\times10^5$~Bq) and $^{137}\text{Cs}$ ($4.07\times10^4$~Bq) sources. Measurements were conducted for 600~s with an integration time window of $2~\mu$s, under 850 V operating voltage of PMT. The energy resolution is directly calculated from the FWHM, and the results are listed in \cref{tab:reflector}.

\subsection{Comparison and Discussion}\label{sec:discussion}

It can be seen from \cref{tab:reflector} that Teflon provides a better energy resolution compared to other materials. This indicates that the properties of reflector material have a significant impact on the energy resolution. Ref.~\cite{Janecek2012} measured the reflectivity of commonly used reflectors.
For the tested reflectors in \cref{tab:reflector}, the reflectivity values are 0.99, 0.985, and 0.97, respectively.
It can be inferred that with the Teflon reflector, the PMT collects more photons due to its higher reflectivity leading to improved energy resolution. The energy resolution with glued ESR in the earlier experiment was worse compared to the resolution with non-glued ESR obtained later.
This is due to the absorption or refraction of light as it passes through the glue.
Moreover, the air gap between the non-glued reflectors and the crystal enhances the collection efficiency of scintillation photons through total internal reflection.

\begin{figure}[!t]
    \centering
    \subfloat[$^{22}\text{Na}$ source.]{\includegraphics[width=0.48\linewidth]{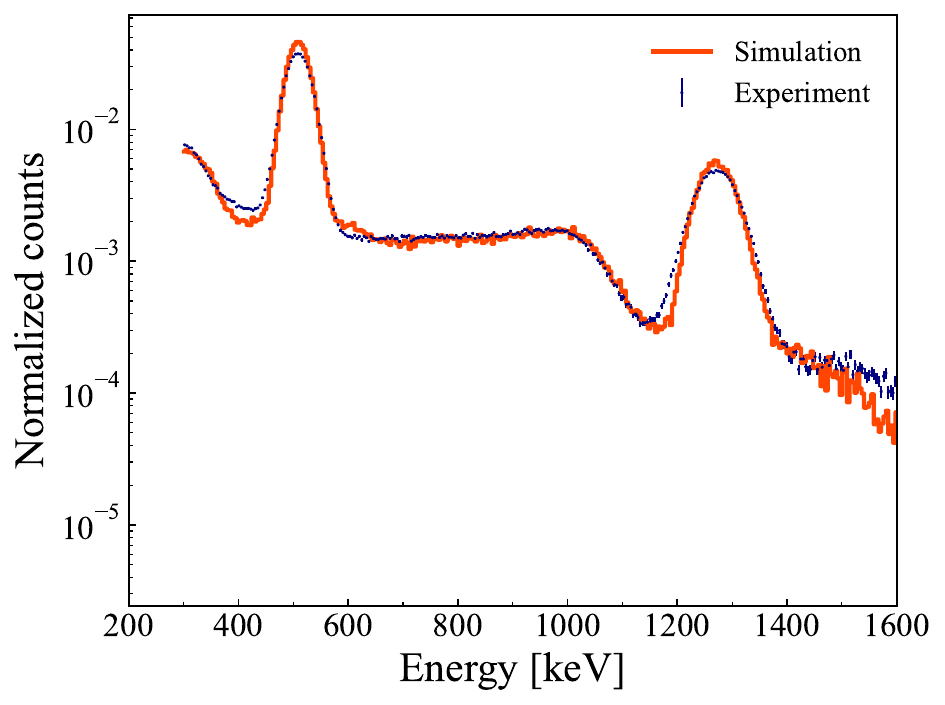}}
    \subfloat[$^{137}\text{Cs}$ source.]{\includegraphics[width=0.48\linewidth]{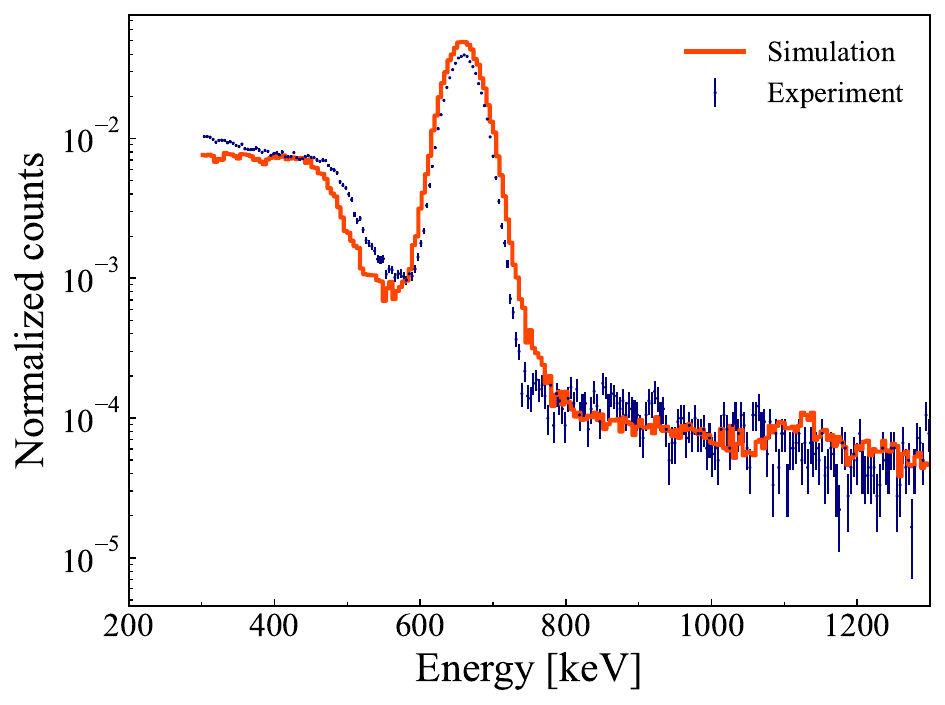}}
    \caption{The energy spectra of $^{22}\text{Na}$ and $^{137}\text{Cs}$ in the experiment (\textit{dots with error bars}) and the simulation (\textit{red line}).}
    \label{fig:compare}
\end{figure}

This study primarily relies on \textsc{Geant4} simulation for the design of the MACE calorimeter, requiring verification of the reliability of various physical processes in the simulation and a comparison with experimental results.
To test the reliability of the simulation results discussed in \cref{sec:2}, we conducted a detailed simulation of the benchmark CsI(Tl) detector using \textsc{Geant4} with the same physics processes.
The CsI(Tl) crystal, Teflon reflector, optical grease, and CR284 PMT volume are modeled identically to the experimental setup. The simulation takes into account the natural background contributions from $^{238}$U and $^{40}$K.
\cref{fig:compare} shows the spectra of calibrated experimental data and those recorded in simulation using $^{22}\text{Na}$ and $^{137}\text{Cs}$ source. The energy resolution is 9.7\% at 511 keV in the simulation, compared to 10.3\% in the experiment (see \cref{tab:reflector}), resulting in a relative difference of 6\%. Note that the difference in the width and height of the full-energy peak between the MC and the data originates from the discrepancy in energy resolution. Other minor discrepancies in the spectra shape are mainly likely caused by the insufficient modeling of electronic noise and nonlinearity, natural backgrounds, materials around the detector, and the simplified optical processes in \textsc{Geant4}~\cite{Alexander2023}. This work focuses on a bench test of the CsI(Tl) crystal. Therefore, important factors mentioned above that are difficult to simulate will be further studied in future prototype development efforts.

In addition, we tested the SiPM readout using the NDL EQR15 11-6060D-S ($6 \times 6~\text{mm}^2$) with a bias voltage set to 42 V in the Teflon reflector case.
Using the same approach, an energy resolution of 26.1\% is obtained. This significantly worse performance is attributed to the much smaller sensitive area of a single SiPM compared to the PMT.
Simulation with the same configuration indicates that using a $4 \times 4$ SiPM array with a total area of $24 \times 24~\text{mm}^2$ could approximately double the number of collected photoelectrons, resulting in an energy resolution of 6.58\% at 511 keV.

The comparison between the experiment and the simulation of the benchmark detector gives the expected results of energy resolution at 511 keV, which validates the design of the MACE ECAL and provides a foundation for further research. The SiPM array readout, with its high PDE, low voltage requirements, and tolerance to magnetic fields, represents a promising upgrade option for the future.

\section{Conclusions and Perspectives} \label{sec:4}
After more than two decades of development of high-intensity muon source and detector technique, the first proposed muon-CLFV experiment in China, MACE, is expected to enhance the sensitivity of M-to-$\bar{\text{M}}$ conversion by more than two orders of magnitude.
This study focuses on the offline-software-based design of the $4\pi$ ECAL geometry for MACE.
The geometry parameters were optimized using the Monte-Carlo simulation method, and the performance such as energy resolution, signal efficiency, and background level were evaluated.
Simulation results indicate an energy resolution of 10.8\% at 511 keV, and a signal efficiency of 78.3\% for annihilation $\gamma$-ray events, greatly outperforming the MACS calorimeter design.
This research establishes a foundation for the subsequent technical development of the ECAL and supports the conceptual design of MACE.

A prototype of ECAL will be constructed for additional validation in progress in these areas.
Further studies are anticipated to evaluate the performance enhancement of ECAL with large-area SiPM array readout.
Upgrades of ECAL are expected with the integration of an inner tracker system.
Moreover, future works will also focus on developing systems for cosmic-ray vetoing, electronic readout, trigger logic, and reconstruction algorithms.

\section{Acknowledgements}
The authors would like to acknowledge the MACE Working Group for their invaluable teamwork. We also extend our gratitude to the colleagues at the SMOOTH lab, as well as to Ke Gong and Chenfeng Yang at Sun Yat-sen University, for their fruitful discussions and assistance related to this work. The simulation benefited greatly from the provision of computing resources by the National Supercomputer Center in Guangzhou. We are also grateful to the Southern Center for Nuclear-Science Theory (SCNT) at the Institute of Modern Physics in the Chinese Academy of Sciences for hospitality.

\bibliographystyle{apsrev4-2}
\bibliography{reference}

\end{document}